\documentclass{article}
\pdfoutput=1
\usepackage{amsmath}
\usepackage{graphicx}
\usepackage{fullpage}

\begin{document}
\title{PETRA IV study with non-interleaved sextupole scheme}
\author{Hung-Chun Chao, Xavier Nuel Gavald\`{a}, and Reinhard Brinkmann, \\DESY, Hamburg, Germany }
\maketitle
%\begin{abstract}
%\end{abstract}

\section{Introduction}
Pioneered by MAXIV's multi-bend lattice\cite{MAXIV}, the trend of synchrotron light source community is to pursue diffraction-limit storage rings.
DESY also plans an upgrade project of PETRA III toward ultra-low emittance PETRA IV.
One of the challenges is the physical constraints such as the DESY's site plan and the existing PETRA tunnel.

Historically, the current tunnel is inherited from an e+e- collider PETRA for particle research in late 70s.
It accommodates a 2.3-km circular accelerator with eight 201.6-m arcs and eight very long straights (108 m and 64.8 m alternately).
Later PETRA has been converted to PETRA II, a pre-accelerator for the lepton-hadron collider HERA.

In 2009, it has been reconfigured to a 6-GeV synchrotron light source PETRA III.
The northeast-to-east octant was adapted from a FODO to a DBA lattice to accommodate 14 undulators for beamlines in Max von Laue Hall.
Twelve RF cavities are clustered in the south straight while the north and west straights house 20 damping wigglers which bring the emittance down to 1 nm.
%The symmetry of the lattice is therefore destroyed.
Hereafter this lattice has only a onefold symmetry.

\begin{figure}[!htb]
   \vspace*{-.5\baselineskip}
   \centering
   \includegraphics*[width=0.5\linewidth]{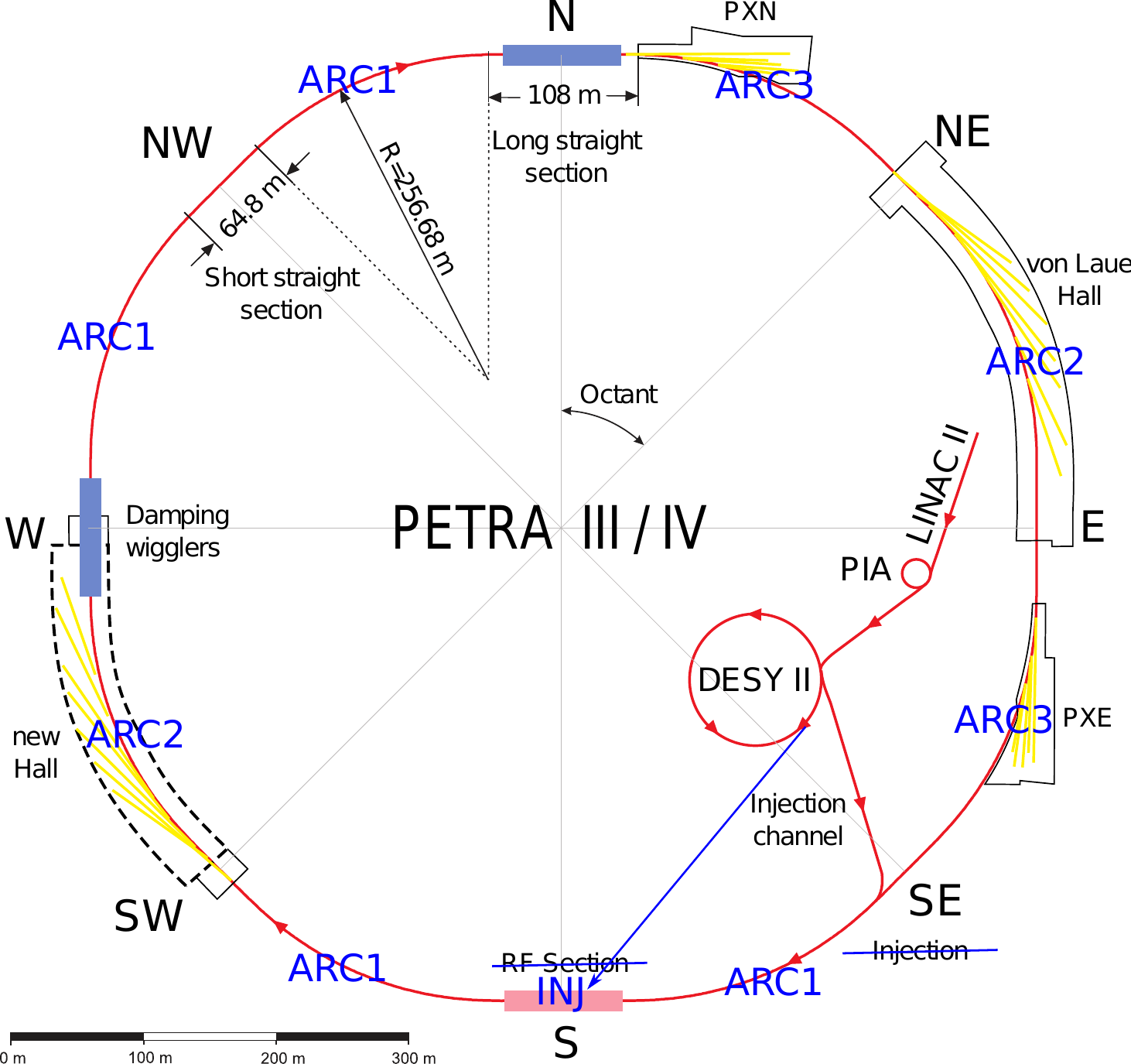}
   \caption{The schematic layout of PETRA III/IV with the new proposed experimental hall. The blue labels indicate the purposed arrangement for PETRA IV.}
   \label{LAYOUT}
   \vspace*{-.5\baselineskip}
\end{figure}

%Afterwards
In 2014-15 the lattice was modified again to change parts of the FODO cells to DBA cells, in order to accommodate 12 more beamlines in 
Paul Peter Ewald and Ada Yonath experimental halls (PXN and PXE).
%the two extension experimental halls PXN and PXE.
This structure then became the standard lattice nowadays.
The schematic layout is shown in Figure~\ref{LAYOUT}.

\section{Design Consideration}
In this study, we are investigating the feasibility to design an upgrade storage ring for PETRA IV by using a non-interleaved sextupole scheme.
%How far can we go with non-interleaved sextupole scheme?
The design goals are set as the following:
\begin{enumerate}
\item the emittance at 6 GeV is in the range of 10-30 pm-rad
(overall effects including damping wigglers, intra beam scattering (IBS), and full coupling operation, etc),
\item at least 5-m straight sections for canted/non-canted IDs,
\item low beta functions to match the electron beam to the photon beam for optimum brilliance,
\item a new experimental hall in southwest-to-west octant.
\end{enumerate}
In the meanwhile, although challenging, we hope to keep:
\begin{enumerate}
\item the original tunnel which has very long straights,
\item beamlines in two extension halls,
\item off-axis injection for accumulation in the storage ring.
\end{enumerate}

For the ultra-low emittance ring design, the multi-bend structure is important since the equilibrium emittance is inverse proportional to the bending angles to the third power.
On the other hand, the ultra-low emittance ring design faces the challenge in terms of sextupole strengths.
With more bending magnets, the dispersions in dipoles can be suppressed, which leads to stronger chromatic sextupoles to correct the chromaticity.
A dilemma exists between small emittance and acceptable sextupole strengths.

Applying innovative technologies to find a satisfactory solution within all engineering limitations is artisan work.
For example, ESRF-EBS has introduced a hybrid multi-bend lattice %by mixing DBA and combined-function FODO lattice\cite{ESRF}.  It features 
featuring proper phase advances between dispersion bumps for the efficient chromaticity correction and the dynamic aperture (DA) optimization\cite{ESRF}.
Other ideas such as longitudinally varying bending or reverse bending push further improvement on the emittance\cite{longu,reverse}.

Thanks to the long arcs in PETRA tunnel, it is relatively easier to fit more bending magnets.
The other advantage of the PETRA tunnel is its very long straights, wherein the RF modules/damping wigglers/injection components can be clustered together.
Therefore the lattice designer can save more space in the arcs to insert even more bending magnets.
(However, the ID beamlines still need some short straights in the arcs.)
A downside of the PETRA tunnel is that more components are needed in the very long straights.
They increase not only the impedance, but also the source of magnet errors.
Their quadrupole components also contribute to additional chromaticities, which require extra sextupole strengths to compensate.

Considering DESY's current site plan, not all of the arcs can be used for the beamline extraction.
Therefore, this design will not be conventional as other third generation light sources which demand symmetry and straights for IDs.
Since the tunnel was built for a collider, it's very natural to following the collider's logic to design such a ring.
Strategically, one can mix different lattices for different purposes due to the tunnel's long circumference.

\section{Building Blocks}
\subsection{ARC1: DMI Arc}
Sextupoles are used to correct the chromaticity but they cause nonlinear effects to shrink the DA.
It is well known that sandwiching an inverse identity transformation section by a sextupole pair ensures a perfect cancellation of the sextupole nonlinear kicks in phase space.
Any other nonlinear element in between is forbidden.
This technique of non-interleaved sextupole scheme requires more space and is often used in collider lattices\cite{KEKB}, since colliders are typically much more spacious than synchrotron light source.

\begin{figure}[!htb]
   \vspace*{-.5\baselineskip}
   \centering
   \includegraphics*[trim=7 150 210 60,clip,width=0.5\linewidth]{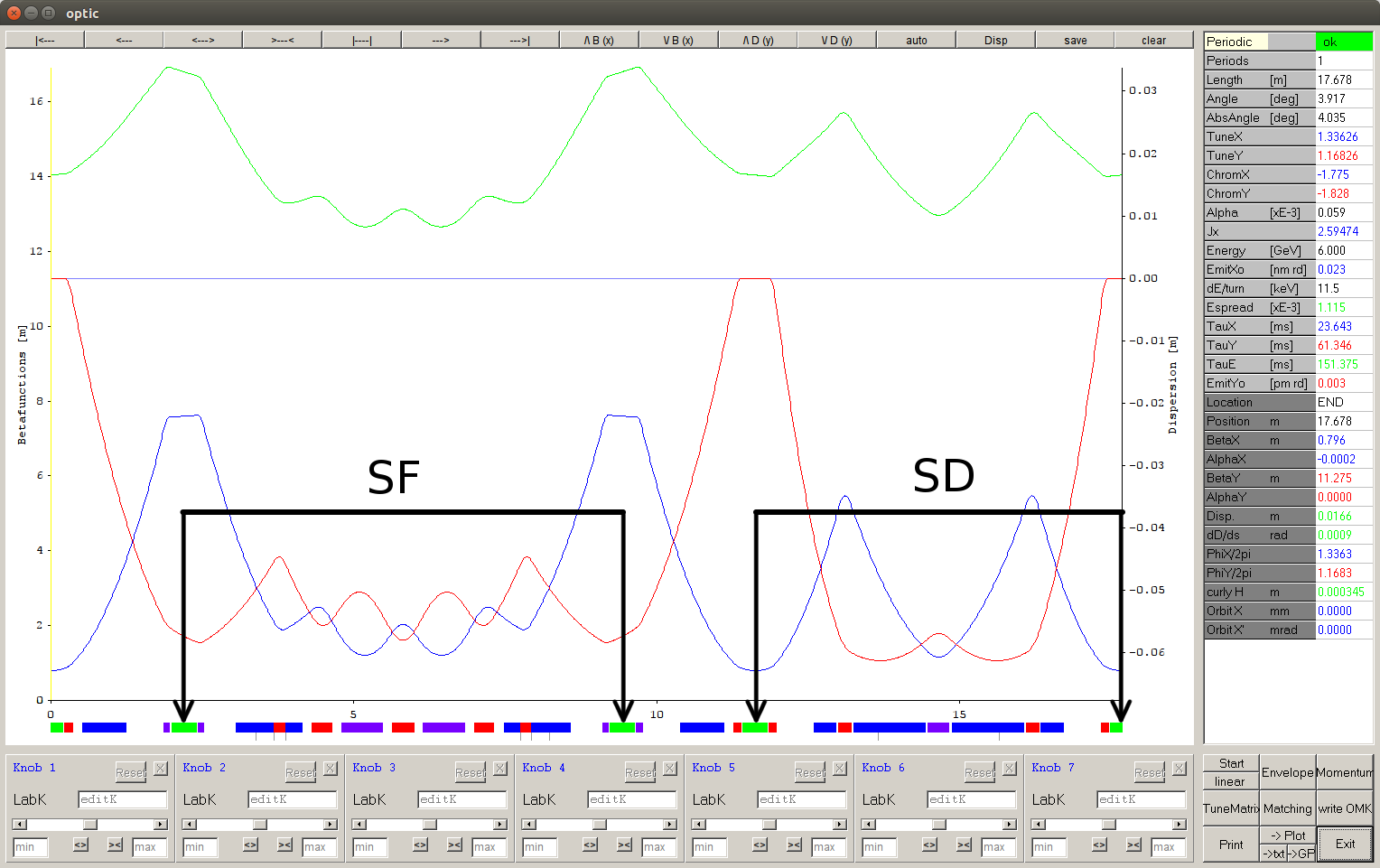}
   \caption{DMI cell.}
   \label{DMI}
   \vspace*{-.5\baselineskip}
\end{figure}

To apply this scheme in the small emittance ring design, one has to create slots for sextupole pairs in a cell with small bending angle and limited length.
%After huge efforts of guided search, finally, two pairs non-interleaved sextupoles are crammed into a 17.7-m long section with the bending angle of 3.92 degree.
Our solution has two pairs of non-interleaved sextupoles crammed into a 17.7-m long arc cell with bending angle of 3.92 degrees.
The emittance of this cell itself is 23 pm-rad.
The lattice structure and its optical functions are shown in Figure~\ref{DMI}.
This type of lattice is named double minus identity (DMI) lattice.

The maximum dipole field is less than 0.26 T, while the maximum quadrupole gradient is limited to 100 T/m.
In some strong quadrupoles there are small reserve bending angles used for better control of the dispersion function.
They can be achieved by slightly shifting the quadrupoles.
The sextupole pairs are located at places with non-zero dispersion and identical optics.
The phase differences are 180 degree, in both directions for each pair.
Due to the high contrast of beta functions, one pair (SF) is more effective for horizontal chromaticity correction and the other (SD) vertical.
%It is the basic building block that the ``unused'' arcs where there is no beamline extracted are filled with.
It is the basic building block for the arcs not accommodating photon beamlines.

Eleven DMI cells make one DMI arc which provides the function of chromaticity correction.
To make the lattice more symmetric and facilitate the optics matching, the lattice is enclosed by an additional section of the part of SD pair in the DMI cell.
In total, there are 11 SF pairs and 12 SD pairs in one DMI arc.

\begin{figure}[!htb]
   \vspace*{-.5\baselineskip}
   \centering
   \includegraphics*[trim=7 150 210 60,clip,width=0.5\linewidth]{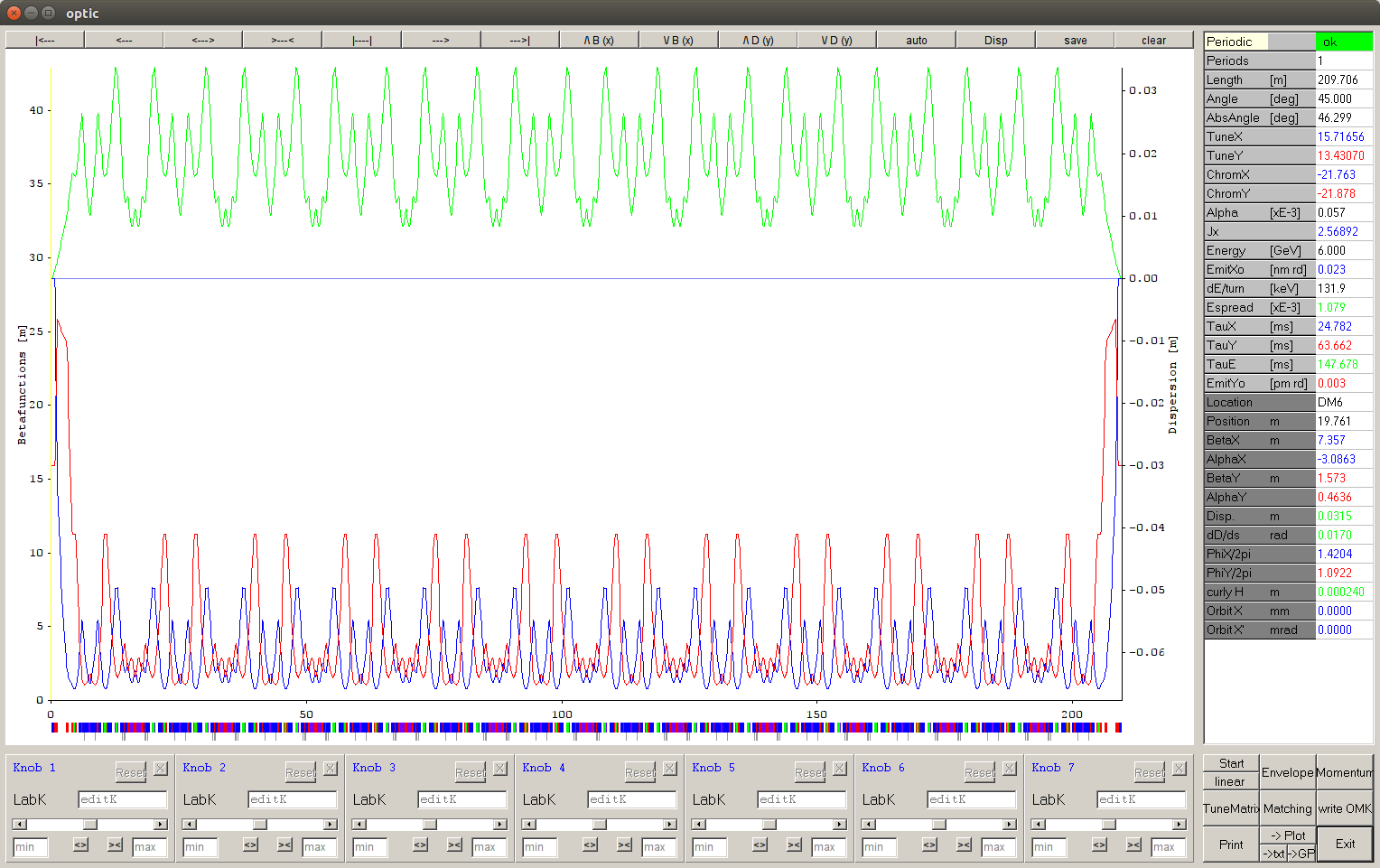}
   \caption{ARC1: DMI arc.}
   \label{ARC-DMI}
   \vspace*{-.5\baselineskip}
\end{figure}

At the ends of the arcs, matching sections to the straight sections are inserted, including dispersion suppression.
The optical functions of DMI arc are shown in Figure~\ref{ARC-DMI}.
%In the ends of the arc, the optics has to be matched to connect to the FODO optics in the straights.
%Because the straights are so long, the dispersion will be greatly amplified if the dispersion function and its slope are non-zero.
%To avoid this, it is necessary to have achromat conditions in the end of the arcs.
%The optical functions of ARC-DMI are shown in Figure~\ref{ARC-DMI}.

\subsection{ARC2: Undulator Arc}
\begin{figure}[!htb]
   \vspace*{-.5\baselineskip}
   \centering
   \includegraphics*[trim=7 150 210 60,clip,width=0.5\linewidth]{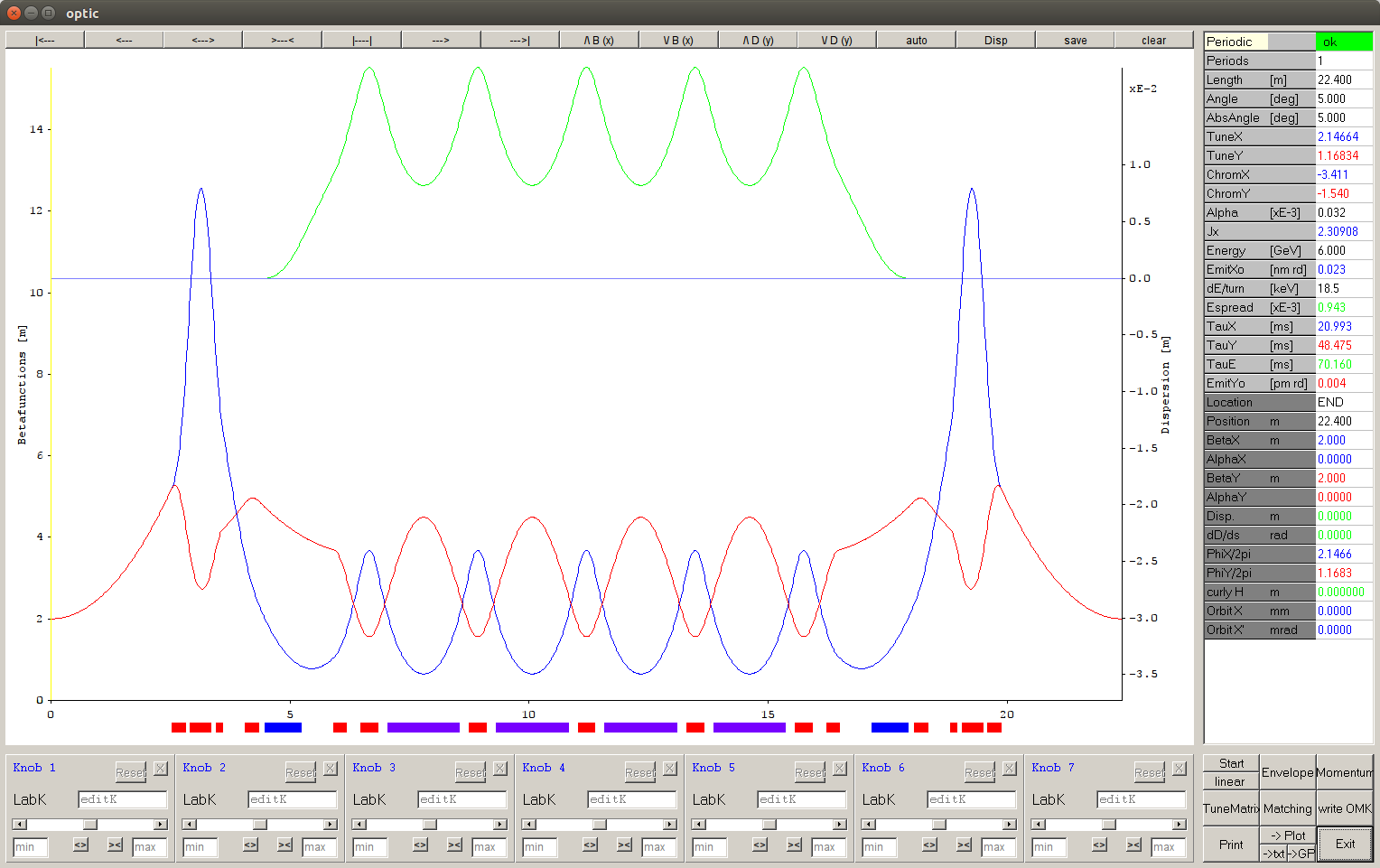}
   \caption{Combined function FODO cell.}
   \label{CFODO}
   \vspace*{-.5\baselineskip}
\end{figure}

The other basic building block is chosen to be the combined-function FODO cell with dispersion suppressors.
The FODO cell yields small natural chromaticity that facilitates the chromaticity correction.
Besides, the gradient component in the bending magnets repartitions the damping coefficients, making the emittance smaller.
The lattice structure and its linear optics functions are shown in Figure~\ref{CFODO}.

Outside of the FODO lattice, the dispersion suppressor is composed of one quadrupole and one bending magnet at proper locations.
The outer quadrupoles flatten and focus the beta functions to small values.
They can offer some tunability for the beta functions in the straights, and also correct the optics when canted/non-canted IDs are presented.

This type of lattice is very suitable for the undulator arcs because it can easily provide an achromat straight longer than 5 m and with low beta functions.
%This type of lattice is a very suitable candidate for the beamline arcs because it can easily provide an achromat straight longer than 5 m and with low beta functions.
%For diffraction limit synchrotron light, the ideal beta functions for straights of length L is roughly $L/\pi$ \cite{controversial}.
The ideal matched beta functions for diffraction limit synchrotron light is controversial\cite{controversial}.
For simplification we choose to control the beta functions in straight centers to be as low as 2 m in both directions. 

%When a canting angle of a few mrad in the straights is presented, the neighboring bending magnet must be lowered.  
%To have less impact on the optics, the bending magnet in the dispersion suppressor is chosen to be a pure function dipole magnet.
%The optics can be recouped via the nearby quadrupoles.

\begin{figure}[!htb]
   \vspace*{-.5\baselineskip}
   \centering
   \includegraphics*[width=0.5\linewidth]{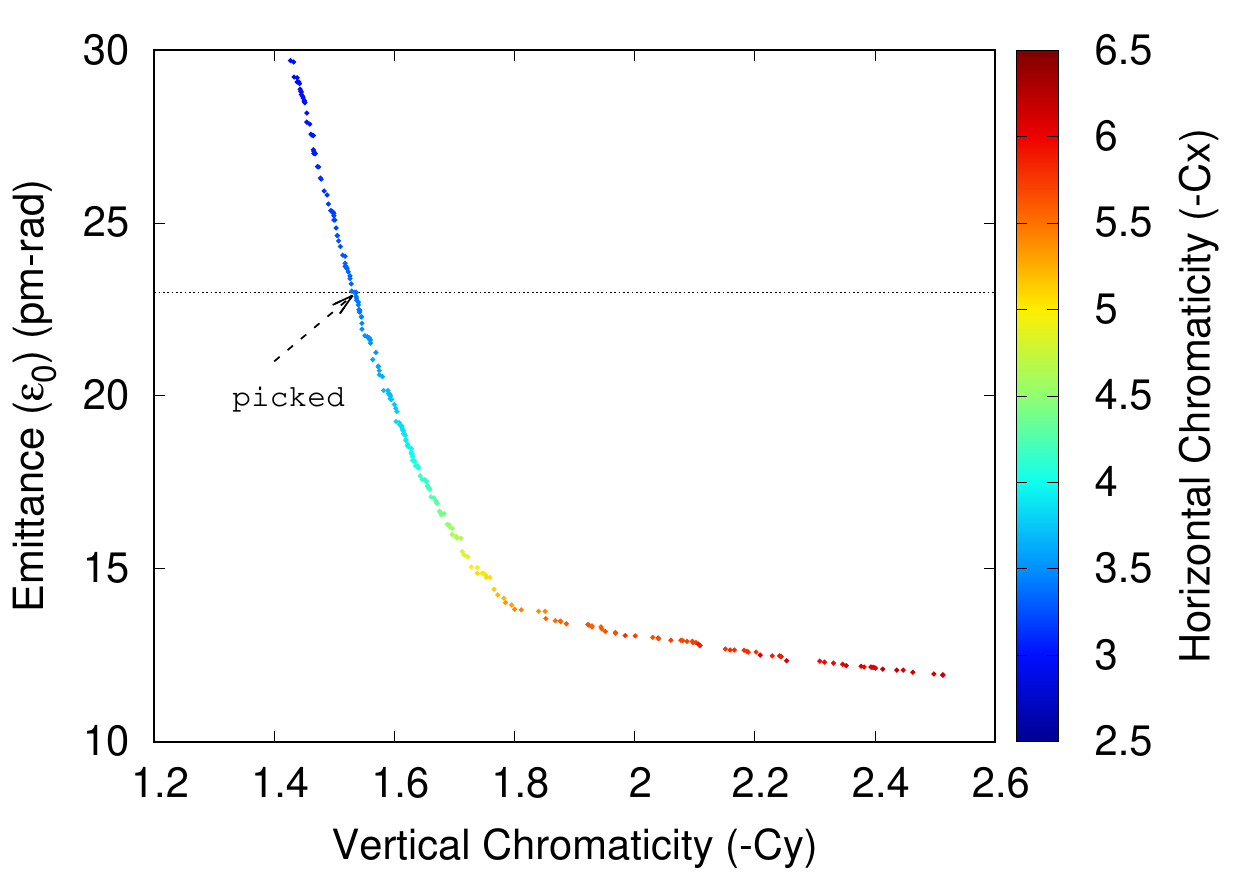}
   \caption{Pareto front of natural emittance vs vertical chromaticity.}
   \label{PARETO}
   \vspace*{-.5\baselineskip}
\end{figure}

In this lattice it's difficult to find ideal slots for non-interleaved sextupole pairs so there is no sextupoles inserted.
Somehow the lattice is still optimized by PMSOEA\cite{IPAC18}.
Instead of the tune spreads involving sextupoles, the objectives can only be quantities related to the linear lattice.
They are the equilibrium emittance and vertical chromaticity since both quantities can not be minimized simultaneously.
The vertical chromaticity was chosen for the consideration of SD strength in DMI cells.
After the pareto front of viable solutions was found, as shown in Figure~\ref{PARETO}, a solution of emittance of 23 pm-rad was picked and plotted in Figure~\ref{CFODO}.
A viable solution is defined as a lattice of this structure which is 22.4 m long and has a straight longer than 5 m with 2-m beta functions in the center.

\begin{figure}[!htb]
   \vspace*{-.5\baselineskip}
   \centering
   \includegraphics*[trim=7 150 210 60,clip,width=0.5\linewidth]{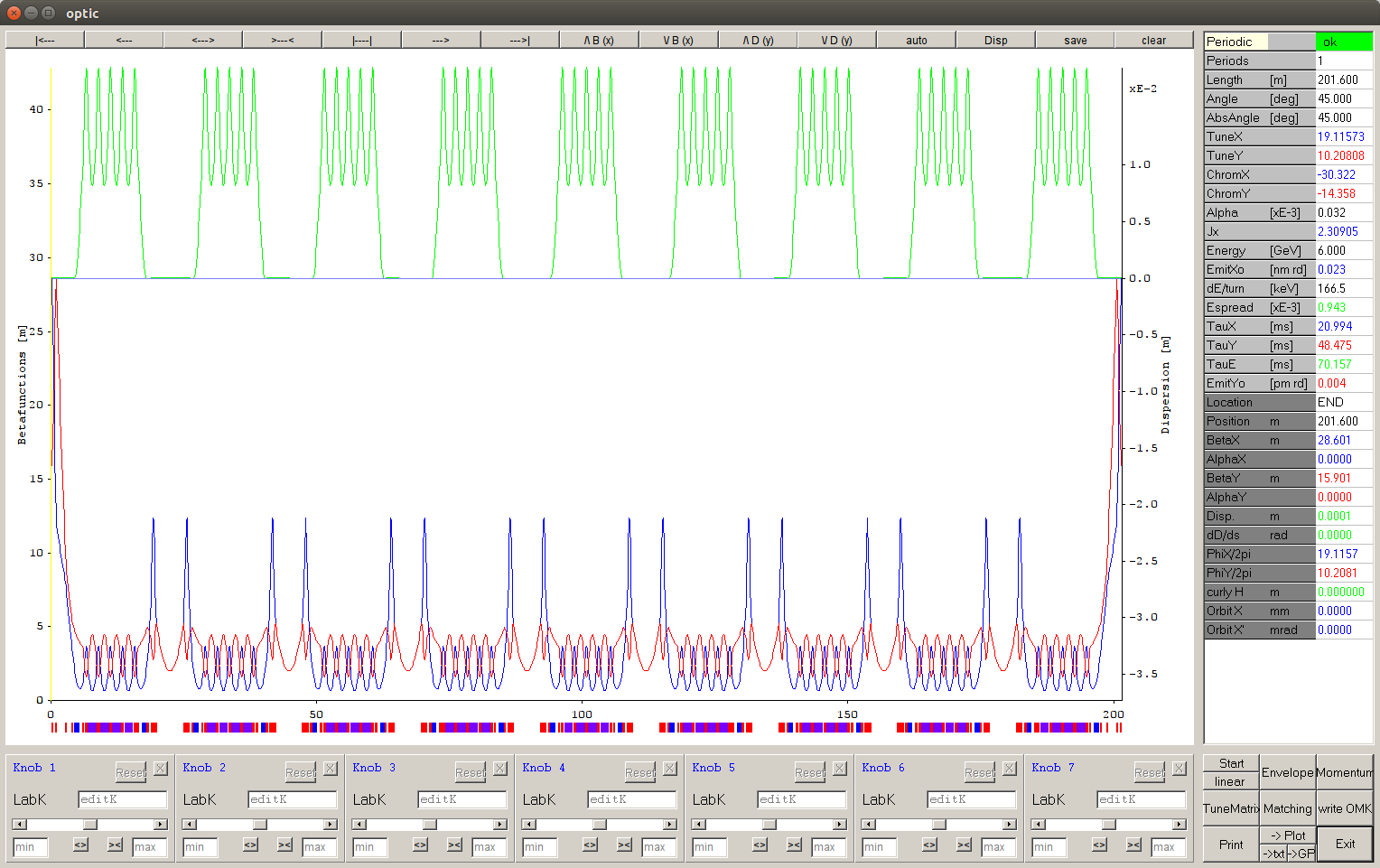}
   \caption{ARC2: Undulator arc.}
   \label{ARC-FODO}
   \vspace*{-.5\baselineskip}
\end{figure}

Nine of this sections make one undulator arc which provides eight straights for undulators.
%As usual, in the ends the achromat condition is needed and the optics are also tuned to match the optics in straights.
An advantage of this design is that the photon beamlines in the existing Max von Laue experimental hall can be kept essentially in its present configuration.
The optical functions of the undulator arc with matching sections are shown in Figure~\ref{ARC-FODO}.

\subsection{ARC3: Hybrid Arc}
\begin{figure}[!htb]
   \vspace*{-.5\baselineskip}
   \centering
   \includegraphics*[trim=7 150 210 60,clip,width=0.5\linewidth]{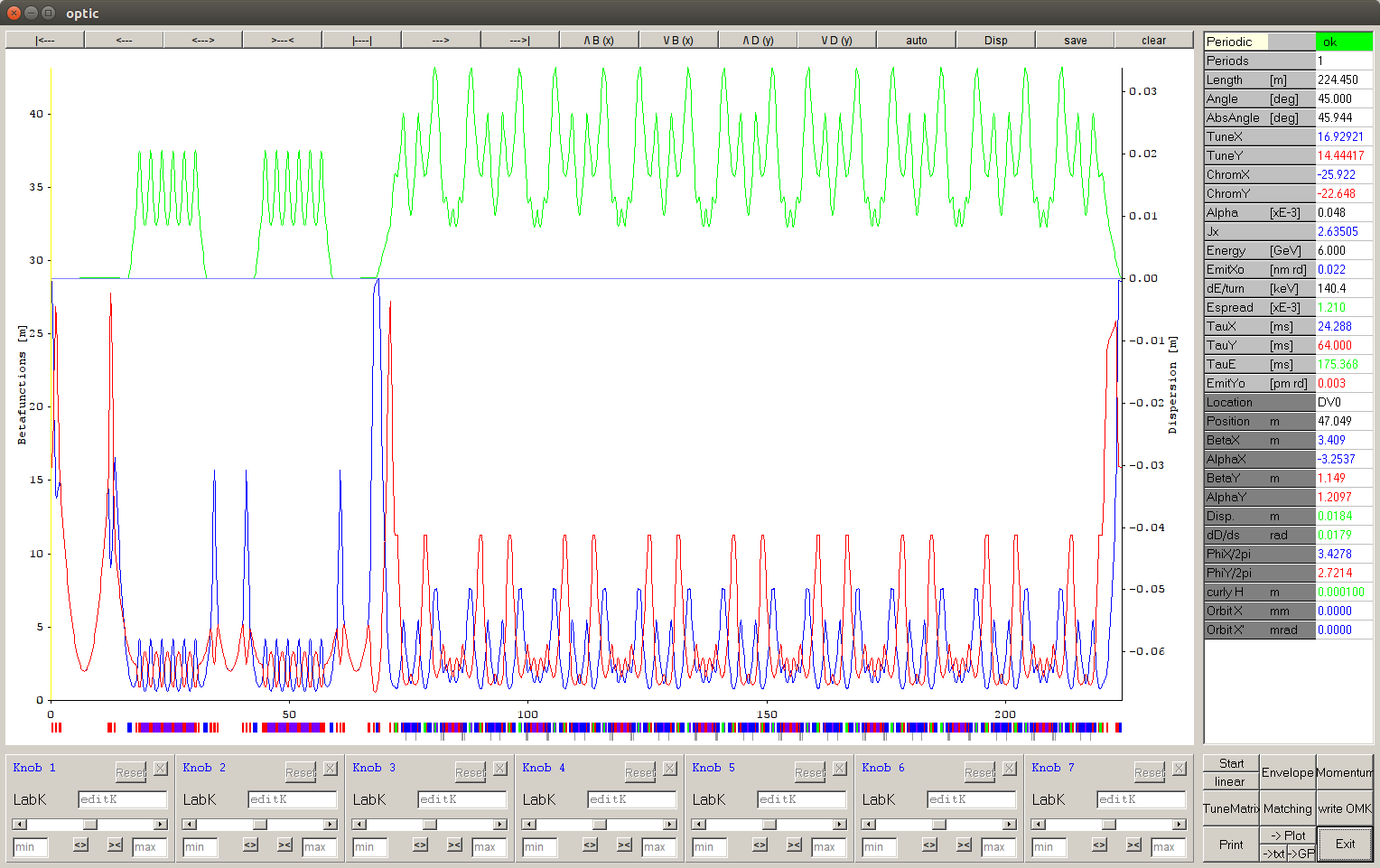}
   \caption{ARC3: Hybrid arc.}
   \label{ARC-Hybrid}
   \vspace*{-.5\baselineskip}
\end{figure}

In PETRA III, ID beamlines in the two extension experimental halls PXN and PXE are extracted in the northeast-to-north and southeast-to-east arcs.
To keep these beamlines, two 5-m straights and one 10-m straight for super ID in the upstream end of each of these two arcs are needed.
The rest of the arcs can be filled with DMI cells.
The resulting lattice is the hybrid arc made by replacing three DMI cells in DMI arc with two custom-made combined function FODO cells and one super ID section, as shown in Figure~\ref{ARC-Hybrid}.
The beta functions are matched to 2 m in the straight centers by the two neighboring quadruple triplets.
For a reference, the length, tunes and chromaticities contributions from all parts are listed in Table~\ref{TAB1}.
%At the upstream end of hybrid arcs a 10-m long straight for super ID is attached.
\begin{table}[!htb]
\small
\centering
\caption{Length, tunes, and chromaticies contributions}
\label{TAB1}
\begin{tabular}{cccccc}
       & Length (m) & $Q_x$ & $Q_y$ & $-C_x$ & $-C_y$ \\
\hline
DMI       & 17.678 & 3.92 & 4.04 & 1.78 & 1.83 \\
\hline
ARC1(4)   & 209.706& 15.7 & 13.4 & 21.8 & 21.9 \\
ARC2(2)   & 201.6  & 19.1 & 10.2 & 30.3 & 14.4 \\
ARC3(2)   & 224.45 & 16.9 & 14.4 & 25.9 & 22.6 \\
Rest      & 613.08 & 4.17 & 4.57 & 4.76 & 4.78 \\
\hline
Total     & 2304 & 139.1 & 107.6 & 204.3 & 166.3 \\
\end{tabular}
\end{table}

\subsection{The Whole Ring}
\begin{figure}[!htb]
   \vspace*{-.5\baselineskip}
   \centering
   \includegraphics*[trim=7 150 210 290,clip,width=0.5\linewidth]{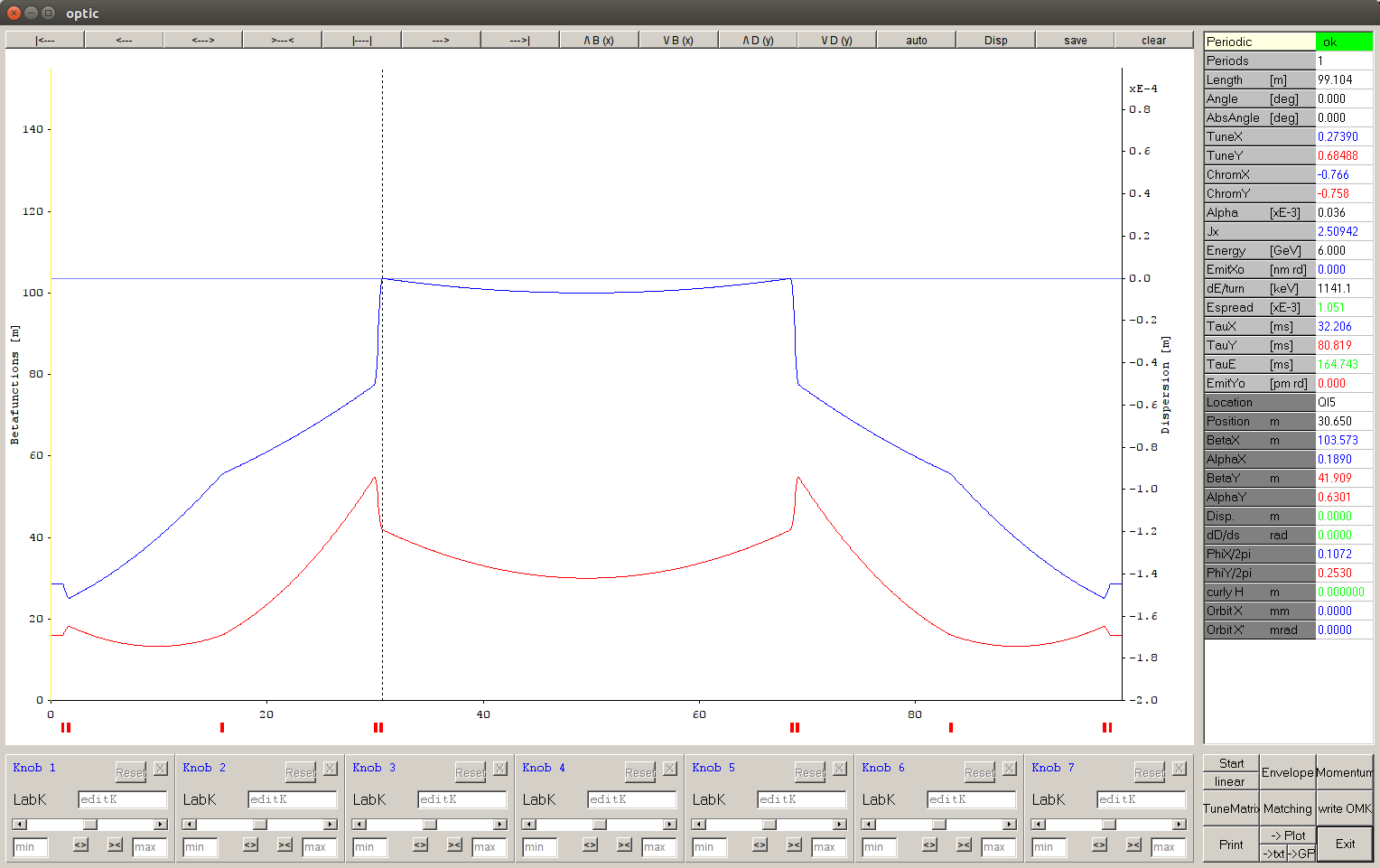}
   \caption{Injection section optics.}
   \label{INJECTION}
   \vspace*{-.5\baselineskip}
\end{figure}

Different types of arcs can be joined by matching the optics in the ends of arcs and in the straights.
The straight lengths also need to be tailored, in order to fulfill the geometry constraints.
The straights in between arcs are inserted with FODO cells and some quadrupoles for the injection.
The optics in the injection section are shown in Figure~\ref{INJECTION}.
Its beta functions are $(\beta_x, \beta_y)=(100,30)$ m in the center of a 37.8 meter long space.
According to Table~\ref{TAB1}, the optics in straights contribute a portion of 2-3\% of total natural chromaticities.

%The straight lengths need to be tailored, in order to fulfill the following geometry constraints:
%\begin{enumerate}
%\item enclose the ideal orbit,
%\item the circumference of 2304 m,
%\item minimize the confliction with tunnel walls.
%\item better alignment with some of the existing beamlines extraction positions.
%\end{enumerate}
%It is another optimization problem.
%For simplification, so far we only considered the first two constraints.
%The number of free parameters is more than enough.

\begin{figure}[!htb]
   \vspace*{-.5\baselineskip}
   \centering
   \includegraphics*[trim=7 150 210 60,clip,width=0.5\linewidth]{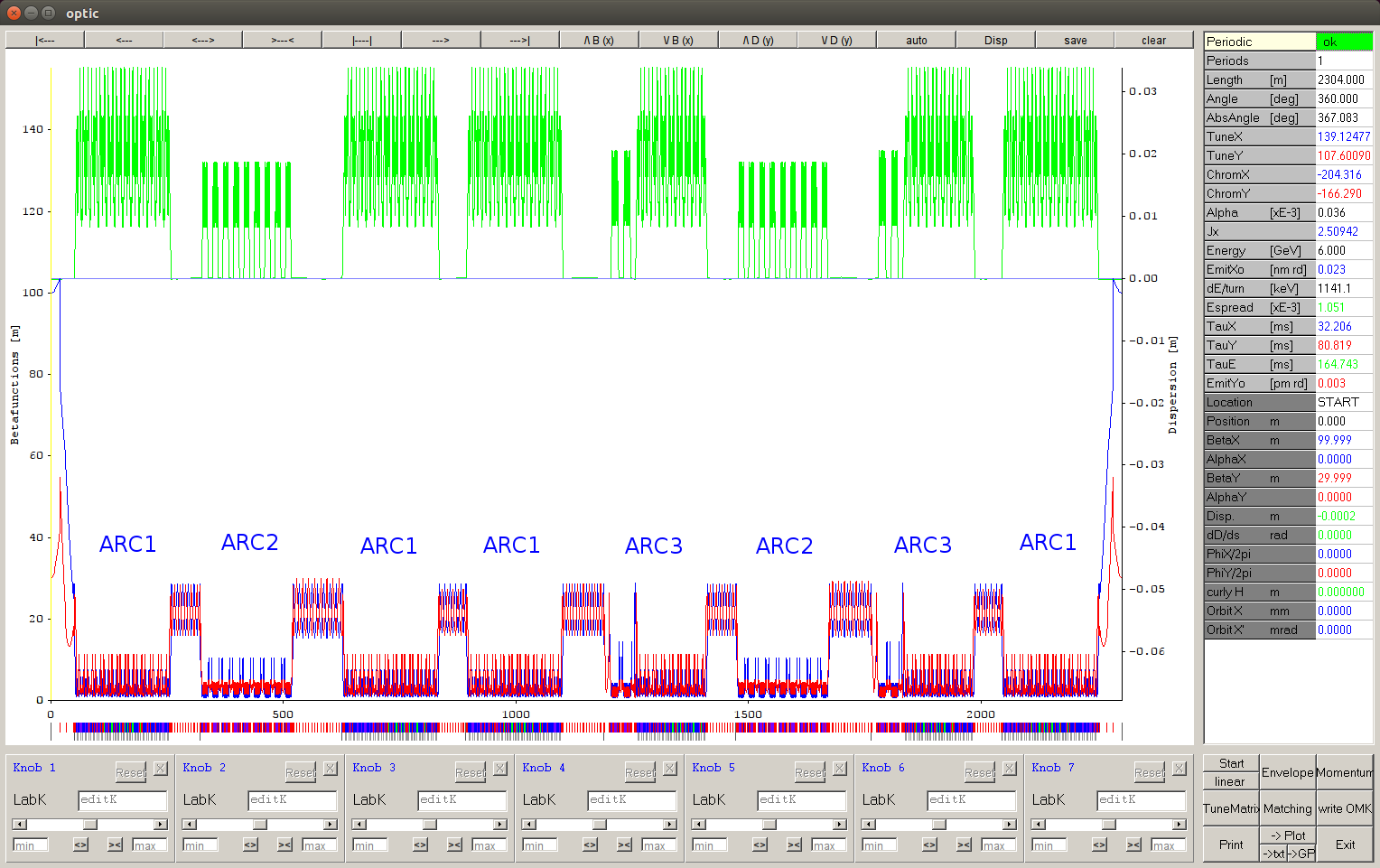}
   \caption{Full ring optics.}
   \label{FULL}
   \vspace*{-.5\baselineskip}
\end{figure}

Following the arrangement of the different arcs depicted in the blue labels in Figure~\ref{LAYOUT} and concatenating all pieces to construct the ring, the full optics along the whole ring shown in Figure~\ref{FULL}.
The parameters are listed in Table~\ref{TAB2}.
The working point needs to be fine tuned via the optics in straights.
\begin{table}[!htb]
\centering
\caption{Storage ring parameters.}
\label{TAB2}
\begin{tabular}{rl}
\hline
Energy   & 6.0 GeV\\
Periodicity     & 1 \\
Circumference   & 2304 m\\
Harmonic Number & 3840 \\
Working Tune & (139.125, 107.601) \\
Natural Chromaticity & (-204.3, -166.3) \\
Horizontal Damping Partition & 2.51 \\
Momentum Compaction & 0.036 $10^{-3}$ \\
Energy Loss & 1.14 MeV \\
Equilibrium Emittance & 23 pm-rad \\
Equilibrium Energy Spread & 1.05 $10^{-3}$ \\
Horizontal Damping Time & 32.2 ms \\
Vertical Damping Time  & 80.8 ms \\
Longitudinal Damping Time & 164.7 ms 
\end{tabular}
\end{table}

Compared to PETRA III, the injection section is moved to south from southeast, while RF modules are moved to other free straights.
In total, there are 132 SDs, 120 SFs, 1542 pure/combined function bending magnets, and around 1300 quadrupoles.
It offers 20$\times$5 m + 2$\times$10 m achromat straights, all with 2-m beta functions in the center.

%Meanwhile, due to the asymmetrical allocations of longer arcs, the geometry are deformed from the original PETRA. 
%The maximum deviation is estimated about 52 cm in one of the long straights.

\section{Nonlinear Dynamics}
In PETRA III there are no sextupoles in DBA sections.
The chromaticity is compensated only by the sextupoles existing in FODOs in arcs.
In this design, the same concept is used.
The non-interleaved sextupole pairs in DMI cells should take care of the compensation of the chromaticities generated all over the ring.
The chromaticity response matrix is 2-by-2 matrix since the optics are identical at SD/SF.
%This can be forseen from the chromaticity response matrix.
It shows that the SF strength is in general weaker than SD.

To compensate the chromaticity generated by DMI cell itself, the required sextupole strengths are only $K_2=(121, -191)$ m$^{-3}$, where $K_2$ is defined as $\frac{1}{B\rho}\frac{\partial B_y}{\partial x^2}$.
When the other parts are included, the required SD strength grows as $K_2=-276$ m$^{-3}$.

Moreover, it is very likely to establish an additional experimental hall in northwest-to-west octant.
In this case one of the ARC1 is replaced by a new ARC2, the required SD strengths will be $K_2=-328$ m$^{-3}$, 19\% stronger than before.
%To alleviate the strength, one can replace the combind function FODO lattice which has no sextupoles with other types of lattice which have sextupoles, such as ESRF-EBS's hybrid MBA with interleaved sextupoles.
%Its chromaticity correction is local and efficient, but the condition of non-interleaved sextupole scheme will be violated.
It is conceivable to alleviate the strength by introducing partial chromaticity correction with interleaved sextupoles in the undulator cells.
That way a compromise may be found between maximizing the DA and the momentum acceptance.

The RF modules will be allocated in achromat sections, so there is no synchro-betatron coupling effects.
A preliminary simulation of 4D tracking without errors shows the DA at the injection point is of the size $\pm 17\times 5$ mm.
%The condition for off-axis injection is fulfilled.
However the corresponding stable range for tune shift with energy is merely up to 1.7\%.
%before tunes hitting half integer 
It is because of the highly asymmetrical structure.

This can be improved by splitting the sextupoles into many families.
Two sextupoles with 180 degree phase difference are still paired together to have the cancellation.
Because the beta functions and the dispersions at SD/SF are identical, the phases play a crucial role.
One way to arrange the families is to distribute the families as even as possible according to the wrapped phases $2\mu_x$ and $2\mu_y$.
The idea behind is to have the internal cancellation of the chromatic half-integer stopband.

Tentatively, the sextupoles are split into 16 families and arranged manually without an optimizing procedure.
Then an evolutionary multi-objectives optimization is applied in order to search the pareto front of on-and-off momentum DAs\cite{MOGA}.
During the optimization the maximum sextupole strength is limited to $|K_2|<300$ m$^{-3}$.
The resulting off-energy stable range is enlarged to 2\% without altering the on-momentum DA too much.

\begin{figure}[!htb]
   \vspace*{-.5\baselineskip}
   \centering
   \includegraphics*[width=0.5\linewidth]{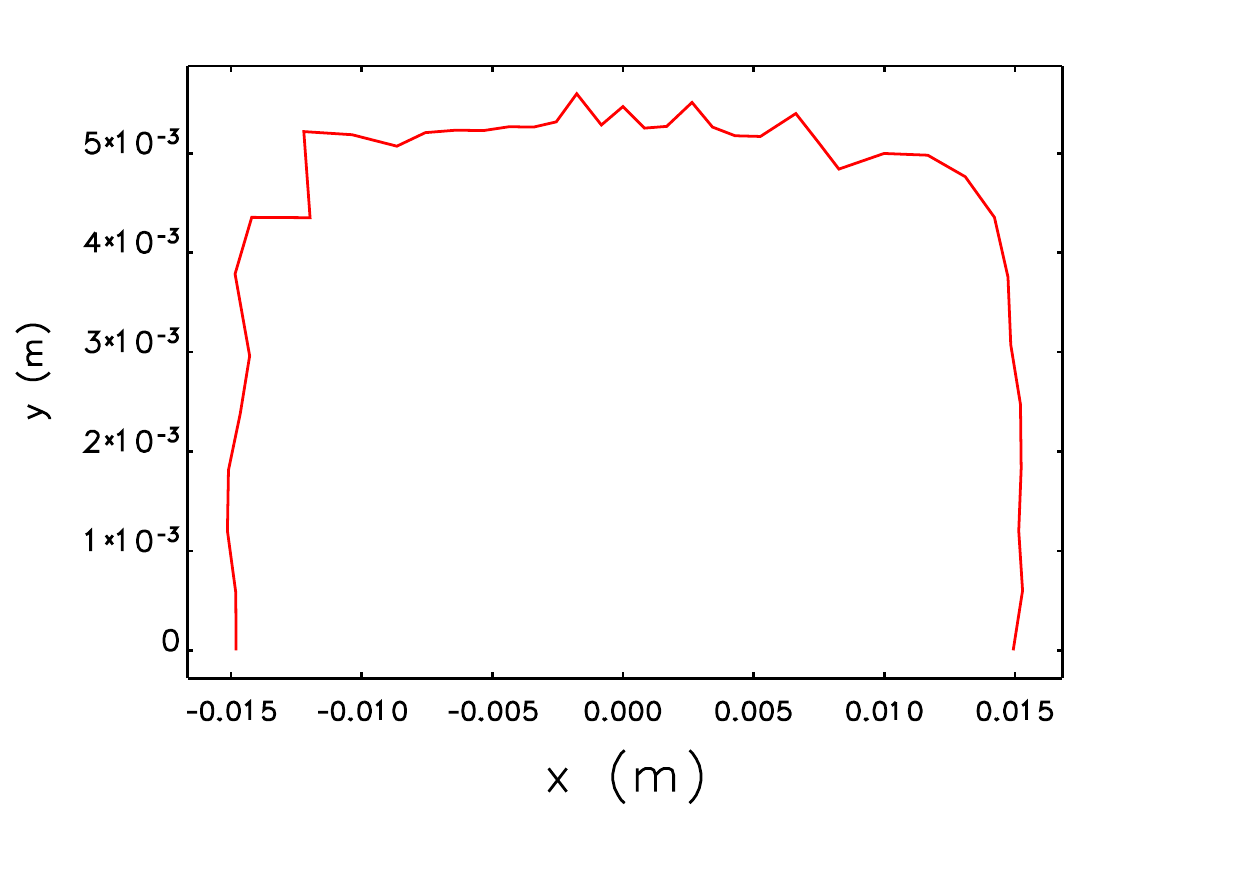}
   \caption{The 1000-turn 6D-tracking DA (without errors).}
   \label{DA}
   \vspace*{-.5\baselineskip}
\end{figure}

The preliminary DA search without errors is carried out by 6D tracking.
The DA at the injection point is shown in Figure~\ref{DA}.
Consequently, the DA is sustainable for the off-axis injection even when it is reduced by a factor of 2 due to imperfections. 
For a more robust simulation with errors, the working point needs to be adjusted to avoid the integer resonance.
%This can be fixed by adjusting the working point.

\begin{figure}[!htb]
   \vspace*{-.5\baselineskip}
   \centering
   \includegraphics*[width=0.5\linewidth]{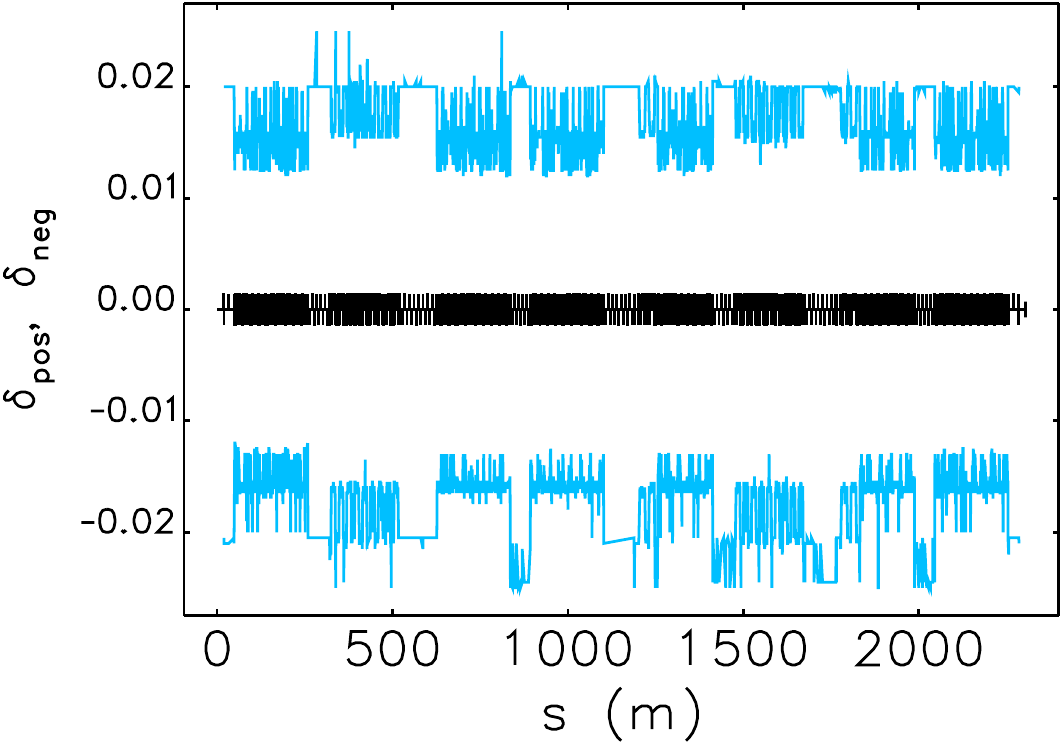}
   %\caption{The LMA along the whole ring, searched by OPA tracking. The resolution is about 10 cm.}
   \caption{The LMA along the whole ring found by 1000-turn and 6D-tracking }
   \label{LMA}
   \vspace*{-.5\baselineskip}
\end{figure}

The local momentum acceptance (LMA) for the Touschek lifetime calculation\cite{Touschek} are shown in Figure~\ref{LMA}.
The values at the entrances of every bending magnet and quadrupole are picked up for averaging.
%Off-axis injection would require a dynamic aperture of about 8 mm, so that with this lattice even when the DA is reduced due to imperfections the accumulation in the ring should be safely possible.
%the bare lattice emittance with 
%The total cavity voltage is assumed 6 MV.
%The physical aperture comes from the radius of inner chamber walls of 13 mm.
%the vertical emittance is assumed 1\% coupled from the bare lattice emittance.

There are two operation modes for users purposed.
The brightness mode bolsters 200 mA in 1600 bunches.
In the timing mode it has 80 mA in 80 bunches.
The estimated Touschek lifetimes for the proposed modes and the effects of IBS and bunch lengthening are listed and compared in Table~\ref{TAB3}.
In this simulation we assume the emittance coupling ratio of 10\%.
The main RF voltage is totally 3 MV, provided by 500 MHz RF cavities.
The bunch lengthening is given by the third harmonic cavities with total voltage as 0.95 MV.
The Touschek lifetimes are increased by the effects of IBS and bunch lengthening because the change density is diluted.
\begin{table}[!htb]
\centering
\caption{Touschek lifetime and parameters.}
\label{TAB3}
\begin{tabular}{ccc}
Operation mode & Brightness & Timing \\
\hline
Total current [mA] & 200 & 80 \\
Number of bunches & 1600 & 80 \\
Charge per bunch [nC] & 0.96 & 7.69 \\
\hline
\multicolumn{3}{c}{Bare lattice}\\
%Emittance [pm-rad] & \multicolumn{2}{c}{22.8} \\
%Energy spread [$10^{-3}$] & \multicolumn{2}{c}{1.05} \\
Bunch length [mm] & \multicolumn{2}{c}{4.34} \\
Touschek lifetime [min] & 21.0 & 2.7\\
\hline
\multicolumn{3}{c}{IBS effect} \\
Emittance [pm-rad] & 38.7 & 66.5 \\
Energy spread [$10^{-3}$] & 1.44 & 1.92 \\
Bunch length [mm] & 5.96 & 7.94 \\
Touschek lifetime [min] & 37.5 & 8.6 \\
\hline
\multicolumn{3}{c}{IBS + 3rd harmonic cavities} \\
%Emittance [pm-rad] & 38.7 & 66.5 \\
%Energy spread [$10^{-3}$] & 1.44 & 1.92 \\
Bunch length [mm] & 28.3 & 30.2 \\
Touschek lifetime [min] & 158.4 & 32.5 \\
\end{tabular}
\end{table}

The emittance blow-up due to IBS can be recouped by damping wigglers but they lead to the reduction of the actual lifetimes.
As a result, the beam accumulation in the storage ring in the brightness mode is possible.
On the other hand, in the timing mode the lifetime is too small for the radiation safety consideration.
%insufficient to accumulate the beam.
%For further improvement, more efforts will be needed to deal with the phases and sextupole families.

%\section{Insertion Device Effect}
%\section{Sensitivity to errors}
%\section{Application}

\section{Conclusion}
In summary, an ultra small emittance storage ring is designed for PETRA IV based on the non-interleaved sextupole scheme and mixing lattices.
The field strengths are reasonable.
The equilibrium natural emittance is 23 pm-rad, without the consideration of damping wigglers and the intra beam scattering effect.
It offers many straights with low and identical beta functions.
The round beam close to the diffraction limit can be delivered with full coupling operation.
The further upgrade to add more beamlines in the potential third experimental hall will need stronger sextupoles.
%The number of beamlines is limited for further upgrade.

The non-interleaved sextupole scheme and the high beta function in the injection straight help the small emittance storage rings design overcome the small dynamic problem, making the off-axis injection possible.
But it comes at the expense of a relatively small LMA due to its asymmetrical structure.
%The capability to accumulate the beam in the storage ring is worrisome.
A remedy is purposed to use more sextupoles families with proper arrangement according to betatron phases.
In a tentative trial the off-energy stable range is extended from 1.7\% to 2\%.
But its capability of beam accumulation is still worrisome.

In an internal discussion in September 2018, this lattice option is not considered as the baseline lattice for the conceptual design report for PETRA IV.
While the lattice design presented here represents a good choice for PETRA IV regarding favorable dynamic properties, other considerations such as a further extendability with more photon beamlines will have an impact on the final design decision.
%The major reason concerns the future upgrade.
%The further study may not be continued.
However, the DMI lattice is still valuable. It is potentially useful in a compact damping ring design. 

\section{Acknowledges}
We wish to thank our colleagues in the PETRA-IV design team for valuable discussions and comments.

\end{document}